\documentclass[11pt]{article}
\usepackage{moriond,epsfig}

\def\Journal#1#2#3#4{{#1} {\bf #2}, #3 (#4)}

\def\NPB{{\em Nucl. Phys.} B}
\def\PLB{{\em Phys. Lett.}  B}

\def\PRD{{\em Phys. Rev.} D}

\def\NPBPS{{\em Nucl. Phys.} B({\em Proc. Suppl.})}
\def\IJMPA{{\em Int. J. Mod. Phys.} A}

\def\be{\begin{equation}}
\def\ee{\end{equation}}
\def\bea{\begin{eqnarray}}
\def\eea{\end{eqnarray}}

\begin{document}
\vspace*{4cm}
\title{LATTICE STUDY OF THE SIMPLIFIED MODEL OF M-THEORY}

\author{P. Bialas and \underline{J. Wosiek} }

\address{Institute of Physics, Jagellonian University, Reymonta 4,\\
Cracow 30-059, Poland}

\maketitle\abstracts{ Lattice discretization of the supersymmetric
Yang-Mills quantum mechanics is reviewed and results of the Monte
Carlo simulations of the simplified model are presented.  The D=4,
N=2, quenched system, studied at finite temperature, reveals
existence of the two distinct regions which may correspond to a
black hole and the elementary 0-branes phases of the M-theory
conjectured in the literature. New results for higher gauge groups
$N < 9 $ lead to the similar picture, however the nature of the
transition between the two phases and its precise scaling with $N$
is yet unresolved. }

\section{Matrix quantum mechanics and its lattice formulation}\label{sec:intro}

Recent developments in nonperturbative string theories have lead
to an exciting hypothesis of the existence of a single theory
(M-theory) which encompasses five known superstring theories and
eleven dimensional supergravity in a unified scheme (for a recent review see e.g. \cite{SEN}).
Even though
its details are not known, the M-theory has a tantalizing
potential to unify all interactions and particles. In particular
it may offer a topological explanation of such fundamental
features
 as three families and fractional charges. It may lead to a standard model gauge group
with $\cal{N}$=1 supersymmetry. It provides understanding of the
Bekenstein-Hawking entropy puzzle and much more. Moreover, Banks,
Fishler, Susskind and Shenker \cite{BFSS} suggested that the
spectrum of M-theory is equivalent to that of a supersymmetric
Yang-Mills quantum mechanics (SYMQM) which results from the
dimensional reduction of the 10 dimensional supersymmetric
Yang-Mills theory. This allows to use a host of nonperturbative
methods to quantitatively study both systems. Accordingly we have
constructed the Wilson discretization of the above quantum
mechanics and proposed to investigate it with the standard lattice
techniques \cite{JW}. The ultimate goal is to study D=10 SYMQM for
the large size of the SU(N) matrices. However, even for this
relatively simple, one dimensional quantum mechanical system this
is a formidable task, the main difficulty being the complex
fermionic determinant (and pfaffian) in D=10, and time consuming
procedures for simulating nonabelian systems of large matrices. On
the other hand the system can be simulated at present in other
regions of the parameter space ($D,N,N_f$) and such a study may
provide some information about its general structure. Therefore we
have decided to set up a systematic lattice study of SYMQM
beginning with the simplest case of $D=4, N=2, N_f=0$ and
gradually extending it as far as possible towards the BFSS limit.
In fact in this talk I will also report on some results for higher
N.
     Supersymmetric Yang-Mills Quantum Mechanics has been intensively studied \cite{UPP}.
      Although
its exact solution is still not available, many results are known and a) can be tested and b)
provide us a guidance from a "simple corner" of parameter space to the ultimate "BFSS corner".

       The action of the SYMQM reads
\begin{equation}
 S=\int dt \left({1\over 2} \mbox{\rm Tr} F_{\mu\nu}(t)^2
                     +\bar\Psi^a(t){\cal D}\Psi^a(t) \right). \label{QM}
\end{equation}
where $\mu,\nu=1\dots D$, all fields are independent of the space coordinates $\vec{x}$
 and supersymmetric fermionic partners
 belong to the addjoint representation of SU(N). The discretized system is put
on a $D$ dimensional hypercubic lattice
$N_1\times\dots\times N_D$ reduced in all space directions to $N_i=1$,
$i=1\dots D-1$.
Gauge and fermionic variables are assigned to links and sites of the new
elongated lattice in the standard manner. The gauge part of the action
reads
\be
S_G=-\beta
\sum_{m=1}^{N_t} \sum_{\mu>\nu}
{1\over N} Re( \mbox{\rm Tr} \, U_{\mu\nu}(m) ),
\label{SG}
\ee
with
$
\beta=2N/a^3 g^2,  \label{beta}
$
and $U_{\mu\nu}(m)= U_{\nu}^{\dagger}(m)U_{\mu}^{\dagger}(m+\nu)
     U_{\nu}(m+\mu)U_{\mu}(m) $,
$U_{\mu}(m)=\exp{(iagA_{\mu}(a m))}$, where $a$ denotes the
lattice constant and $g$ is the gauge coupling in one dimension.
The integer time coordinate along the lattice is $m$. Periodic
boundary conditions $U_{\mu}(m+\nu)=U_{\mu}(m)$, $\nu=1\ldots
D-1$, guarantee that Wilson plaquettes $U_{\mu\nu}$ tend, in the
classical continuum limit, to the appropriate components
$F_{\mu\nu}$ with space derivatives absent. In this formulation
the projection on gauge invariant states is naturally implemented.
Eq. (\ref{SG}) is the basis of the MC simulations of the still
more simplified, D=4, N=2 and $N_f=0$ (quenched), model \cite{JW}.
Some preliminary results for higher gauge groups $N<9$ are now
also available and will be shortly discussed.
\section{Results}
As a first problem we have chosen the question of the phase
structure of the model. A very interesting feature of the M-theory
is the solution of the Bekenstein-Hawking entropy puzzle in terms
of the elementary branes \cite{STRO}. In particular, the theory
predicts existence of the phase transition between a low
temperature "black hole" phase and a high temperature phase
described by the elementary branes \cite{MAR} \footnote{ The full
phase structure of the M-theory is expected to be much more
complex {\em op. cit.}}. Threfore a natural question arises if the
simplified model (i.e. SYMQM) posesses any nontrivial phase
structure \cite{KAB}. At the same time it is well established that
QCD (or pure Yang-Mills theory)
 has two different phases (e.g. confinement and deconfinement).
  Since the action (\ref{SG}) is basically QCD-like, one
 may expect that the dimensionally reduced model
 may indeed exhibit similar phenomenon. To check this we have measured the distribution
of the trace of the Polyakov line
\be
    P={1\over N}  \mbox{\rm Tr} \left( \prod_{m=1}^{N_t} U_D(m)
\right).  \label{poly}
\ee
which is a very sensitive determinant of the two phases in QCD. Similarly to lattice pure gauge
system,
symmetric concentration of the trace around zero indicates the low temperature phase
with $<P>=0$ (here a "black hole" phase) while clustering around $\pm 1$ (or, for arbitrary N,
 around the elements of $Z_N$) is characteristic of the high temperature phase
(here the elementary 0-brane phase).

 There is one important difference between the dimensionally reduced
 model and the pure gauge system in the large spatial volume.
Of course the one dimensional system with local interactions cannot
have a phase transition for finite N. However at infinite N the sharp singularity may occur.
This suggests that the number of colours plays a role of a volume (indeed the number of degrees of freedom
is proportional to $N^2$). Consequently, and similarly to a small volume systems,
 we expect to see some signatures of a phase change for finite
 and even small N . However the generic singularities
would develop only at infinite N. In fact studying the N dependence of the above signatures
allows often to determine the properties of the infinite N phase transition via the final size
 scaling analysis.

In Ref.\cite{JW} above program was carried out for D=4, N=2 and in
the quenched approximation $N_f=0$. It was found that the
distribution of (\ref{poly}) is indeed changing from a convex to a
concave one at some finite value of $\beta=\beta_c(N_t)$. Moreover
the dependence of $\beta_c$ on $N_t$ turned out to
 be consistent with the canonical scaling. The best fit to Monte Carlo results gave
$\beta_c(N_t)=(0.17\pm 0.05)N_t^{(3.02\pm 0.33)}$ in good
agreement with the expected dependence $\beta_c\sim N_t^3$. This
indicated that in the continuum limit the transition occurs at
{\em finite} temperature $T_c=A (g^2 N)^{(1/3)}$, with $A=.28\pm
0.03$. We have also measured the average size of the system
$R^2=g^2\sum_a (A_i^a)^2$  which is shown in Fig.1 for different
values of the temperature. Our results agree qualitatively with
the mean field calculation at large $N$ \cite{KAB}. Moreover, the
pseudocritical temperature determined from Fig.1 is consistent
with the value of $A$ quoted above, which to our knowledge was
determined for the first time. Obviously there is a long way
between $N=2$ and the BFSS limit $N\rightarrow\infty$. Even though
there is an intriguing evidence from lattice simulations that
large N limit may be reached rather early in N (at least for some
observables) \cite{TEP,WIT}, it is very important to repeat this
calculation for higher N. In particular the coefficient $A$ may
depend on $N$.

\begin{figure}[h]
\parbox{6cm}{Figure 1. Extrapolated to the continuum limit size of the system as a function of
 the temperature (in the units of the one dimensional gauge coupling $g$).}
\ \hspace*{1cm} \
\parbox{8cm}{
\psfig{figure=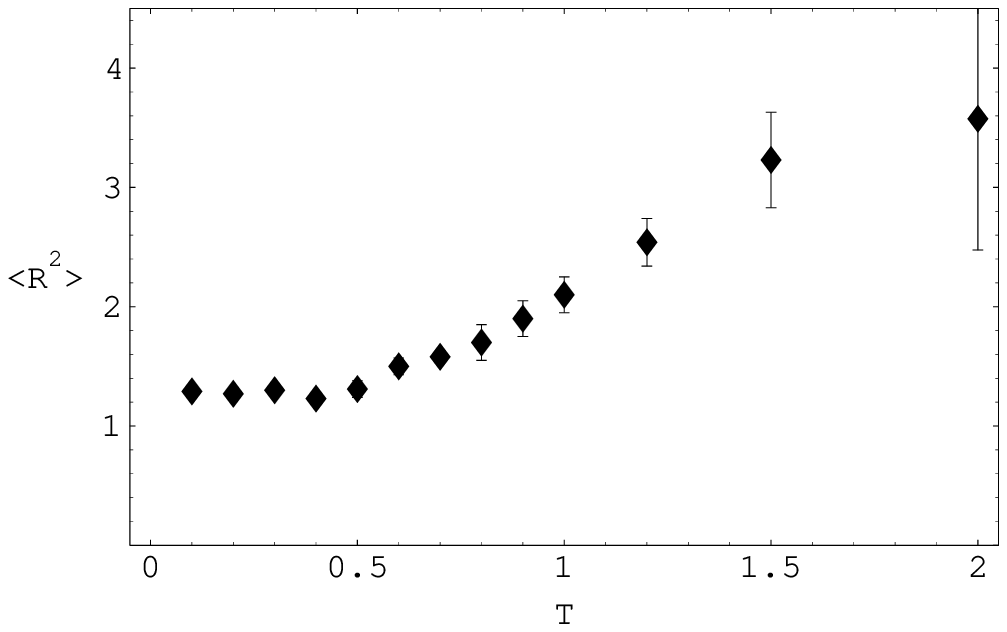,height=4cm} }
\end{figure}

Currently we are extending this analysis to higher gauge groups (and still for $N_f=0$).
Detailed account of our results
will appear elsewhere. Main points are the following. For $N>2$ the trace P
in Eq.(\ref{poly}) is complex. The condition det(U)=1 constraints the trace to lie inside the
"N-star" in the complex plane (see Fig.2). In the low temperature phase the distribution is symmetric and peaked
around $P=0$ ensuring $<P>=0$, while in the high temperature phase one expects P's to be
 concentrated around the elements of $Z_N$. A sample of our preliminary results for N=3,5,8
and at high and low temperatures $(\beta\sim T^3)$
 is shown in Fig.2. Indeed one clearly sees existence of both (high and low temperature)
 regimes. Moreover, in the high temperature region, the system has a tendency to be stuck
in one of the $Z_N$ minima. 
This is a typical indication of the spontaneous symmetry breaking which might occur at
 $N\rightarrow \infty$ leading eventually to a non-zero value of the order parameter $<P>$.
We expect to accumulate soon enough data to be able to decide if the two above regions are
separated by the genuine phase transition or a smooth cross-over.

\begin{figure}[h]
\psfig{figure=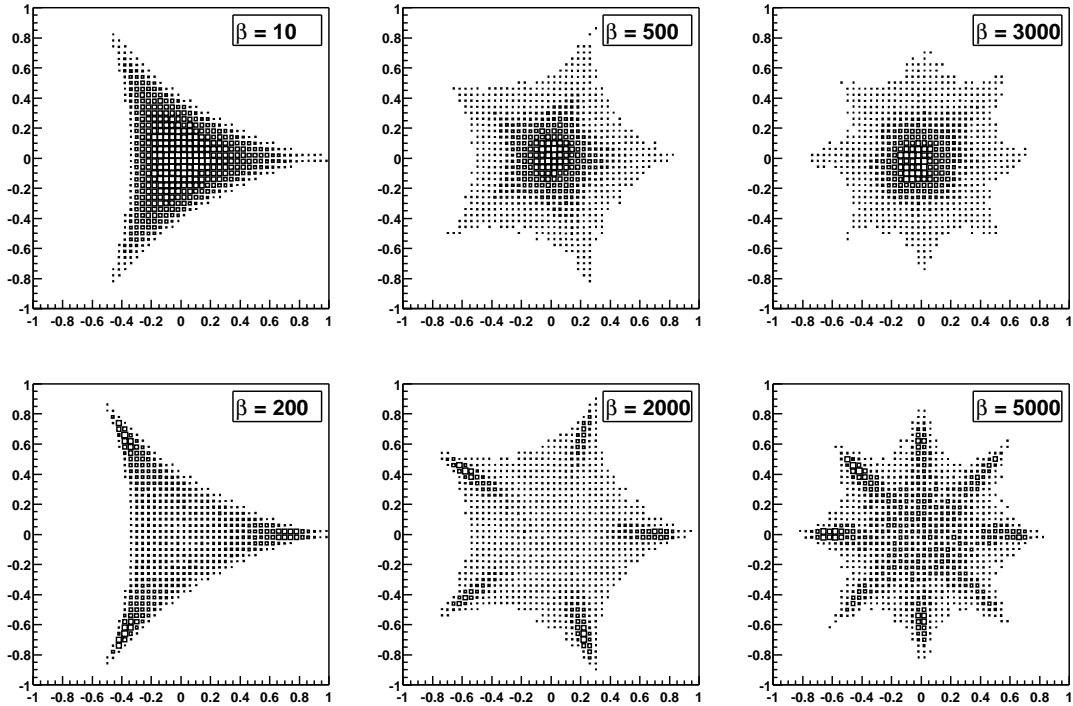,height=10cm}
\caption{Distribution of the Polyakov line in the low (upper row) and high (lower row)
         temperature phases of the quenched Yang-Mills quantum mechanics with $N=3,5,8$. }
\label{fig:radish}
\end{figure}

To conclude, the lattice simulations provide a novel  approach to
the Yang-Mills quantum mechanics and possibly to the M-theory.
Many problems (e.g. identification of the graviton multiplet) can
be formulated and quantitatively attacked  in this framework. On
the other hand much more work is required to continue development
of this fascinating subject.
 In particular, construction of the new algorithms
capable to deal effectively with dynamical fermions propagating in this one dimensional system
is an open question.
\section*{Acknowledgments}
This work is supported by the Polish Committee for Scientific Research
under the grant 2 P03B 019 17.

\end{document}